\documentclass[12pt]{iopart}
\expandafter\let\csname equation*\endcsname\relax
\expandafter\let\csname endequation*\endcsname\relax
\usepackage{amssymb,amsmath,amsfonts,bbold}
\usepackage{graphicx}
\begin{document}
\title[Self-adjoint Dirac type Hamiltonians in one space dimension with a mass jump]{Self-adjoint Dirac type Hamiltonians in one space dimension with a mass jump.}

\author{L. A. Gonz\'{a}lez - D\'{\i}az}

\address{Laboratorio de Din\'amica no Lineal y Sistemas Complejos. Centro de F\'{\i}sica, Instituto Venezolano de Investigaciones Cient\'{\i}ficas, Caracas 1020 - A, Venezuela.}
\ead{lgonzale@ivic.gob.ve}

\author{Alberto A. D\'{\i}az}

\address{Centro de Estudios Avanzados,
Instituto Venezolano de Investigaciones Cient\'{\i}ficas, Caracas 1020 - A, Venezuela.}
\ead{aldiaz@ivic.gob.ve}

\author{S. D\'{\i}az - Sol\'orzano}

\address{Departamento de F\'{\i}sica, Universidad Sim\'on Bol\'{\i}var,
Sartenejas, Edo. Miranda 89000, Venezuela.}
\ead{sttiwuer@yahoo.es}

\author{J. R. Darias}

\address{Laboratorio de \'{O}ptica y Fluidos, Universidad Sim\'on Bol\'{\i}var,
Apartado Postal 89000, Caracas 1080 - A, Venezuela.}
\ead{jrdarias@usb.ve}

\begin{abstract}
Physical self-adjoint extensions and their spectra of the one-dimensional Dirac type Hamiltonian operator in which both the mass and velocity are constant except for a finite jump at one point of the real axis are correctly found. Different boundary conditions on envelope wave functions are studied, and the limiting case of equal masses (with no mass jump) is reviewed. Transport across one-dimensional heterostructures described by the Dirac equation is considered.
\end{abstract}


\section{Introduction}\label{intro}

The discovery of graphene ended the belief that the Dirac equation useless in condensed matter physics \cite{novoselov1,novoselov2}.
The scientific and technological potential for exploiting charge carriers and quasiparticles with relativistic behavior in tunable condensed matter and atomic physics systems is attracting much attention \cite{zhu,ruostekoski,gomes,zhang}. In this regard, an important question, as yet only partly explored, remains whether quasi - one dimensional graphene systems support exclusively Dirac - Weyl massless or constant - mass Dirac fermions, or they can induce relativistic quantum field behaviors that require the consideration of position - dependent mass term \cite{yannouleas}. The mass jump case was firstly consider in \cite{tamarchenko} for the spherical defects in the $III-V$ semiconductors and then in \cite{firsova1,firsova2,ktitorov} for cylindrically symmetric defects in graphene, where the presence of the mass jump was traced back to the graphene sublattice symmetry violation, which is natural but for a short range defect. Using the Dirac - Weyl equation, the break - down of the sublattice symmetry (the equivalence of the two triangular lattices in graphene) can be described in terms of the effective mass that can be position - dependent. The existence of localized Dirac fermions in graphene with inhomogeneous effective mass was considered in \cite{jakubsky}, where the conditions under which the Dirac fermions are confined are found.

Models with an abrupt discontinuity of the mass and velocity at one point can be used for describing the behavior of a quantum particle moving between two different materials, i.e. an electron moving in a media formed up by two different materials. In each material the particle behaves as if it had a different mass and velocity. The discontinuity point represents the junction between these two materials.

We propose a simple model that describes an electron moving in a media formed up by two different materials given by a one - dimensional system in which the mass and velocity are constant except for a finite jump at one point of the real axis, which is chosen to be the origin for simplicity,
\begin{equation}\label{piecemass}
m(x)=\begin{cases}
m_{l} &\textrm{if}\hspace{.2cm} x<0\\
m_{r} &\textrm{if}\hspace{.2cm} x>0,
\end{cases}
\end{equation}
where $m_{l}$ and $m_{r}$ ($m_{l}\neq m_{r}$) are the masses at rest on the left and right, respectively. In this case, the Hamiltonian operator has the functional form
\begin{equation}\label{piecehamiltonian}
H=\begin{cases}
-i v_{l}\sigma_{x}\frac{d}{dx} + m_{l}v^{2}_{l}\sigma_{z} &\textrm{if}\hspace{.2cm} x<0\\*[.2cm]
-i v_{r}\sigma_{x}\frac{d}{dx} + m_{r}v^{2}_{r}\sigma_{z} &\textrm{if}\hspace{.2cm} x>0,
\end{cases}
\end{equation}
where $v_{l},\,v_{r}$ are the Fermi velocities in each medium and
\begin{equation}\label{pauli}
\sigma_{x}=\begin{pmatrix}0&1 \\ 1&0\end{pmatrix},\;\;\sigma_{z}=\begin{pmatrix}1&0 \\ 0&-1\end{pmatrix}
\end{equation}
are the Pauli matrices.

The finding appropriate of the boundary conditions in this kind of models is very important to describe the correct physics. The study of the role of the boundary conditions of quantum systems has became a recent focus of activity in different branches of physics \cite{asorey,asorey2}. Some examples of quantum physical phenomena which are intimately related to boundary conditions are the Casimir effect \cite{casimir}, the role of edge states \cite{john} and the quantization of conductivity in the Hall effect \cite{thouless}.

In this paper, we show that the operator \eqref{piecehamiltonian}, in a suitable domain, has infinite self-adjoint extensions. All the self-adjoint extensions have real discrete spectrum. Thus, all self-adjoint extensions describe bound states only, but not all the extensions are physically acceptable. We examine which extensions could to play an interesting role according to physical arguments.

The paper is organised as follows: In section \ref{extensions}, we find the set of all possible self-adjoint
extensions of $H$. In section \ref{spectra}, we calculate the reflection and transmission coefficients for all self-adjoint extensions, and we use constraints from physical arguments to reduce the set of all possible self-adjoint extensions. From the equation of the poles of the scattering coefficients, we obtain the spectrum that characterizes each self-adjoint extensions.

\section{Self - adjoint extensions of H}\label{extensions}
We will follow to Reed \cite{simon} and  Naimark \cite{naimark} to construct the self-adjoint extensions of $H$. To construct
the self-adjoint extensions of the operator $H$ we must begin by defining the smaller domain where the action operator makes
sense. In this section we will assume that the operator $H$ is densely defined. The domain of the operator $H$, $\mathcal{D}(H)$, is
\begin{equation}\label{domh}
\mathcal{D}(H)=\left\{\psi\in W^{2,1}(\mathbb{R})\otimes\mathbb{C}^2\, ,\,\psi(0^-)=\psi(0^+)=0\right\},
\end{equation}
where $W^{2,1}(\mathbb{R})$ is the corresponding Sobolev space, and $\psi$ is a two - component spinor wave function
\begin{equation}\label{spinor}
\psi(x)=\begin{pmatrix}\psi_{a}(x) \\ \psi_{b}(x)\end{pmatrix}
\end{equation}
The operator $H$ is symmetric and closed. Let $H^{\dag}$ the adjoint of $H$, with domain
\begin{equation}\label{domhdag}
\mathcal{D}(H^\dag)=\left\{\psi\in W^{2,1}(\mathbb{R} \backslash \left\{0\right\})\otimes\mathbb{C}^2\right\}.
\end{equation}
Note that $\mathcal{D}(H)\subset\mathcal{D}(H^{\dag})$. The deficiency subspaces of $H$ are given by
\begin{equation}\label{subdefecto}
\mathcal{N}_{\pm}=
\left\{\psi_\pm\in\mathcal{D}(H^\dag)\,,\, H^\dag\psi_\pm=\pm i\psi_\pm\right\},
\end{equation}
with the respective dimensions $n_{+}$ and $n_{-}$. These are called the deficiency indices of the operator $H$
and will be denoted by the ordered pair $(n_{+},n_{-})$.
The normalized solutions of
$H^\dag\psi_\pm=\pm i\psi_\pm$ are
\begin{subequations}\label{soldefecto}
\begin{eqnarray}
\label{soldefmas-mas}
\psi_+^{(+)}(x)&=&\left[\tfrac{(1+m_r^2 v_r^4)}{v_r^2}\right]^{1/4}\begin{pmatrix}1 \\ i\tfrac{\sqrt{1+m_r^2 v_r^4}}{i+m_r v_r^2}\end{pmatrix}\theta(x)\,e^{-\tfrac{\sqrt{1+m_r^2 v_r^4}}{v_r}x},\\
\label{soldefmas-men}
\psi_+^{(-)}(x)&=&\left[\tfrac{(1+m_l^2 v_l^4)}{v_l^2}\right]^{1/4}\begin{pmatrix}1 \\ -i\tfrac{\sqrt{1+m_l^2 v_l^4}}{i+m_l v_l^2}\end{pmatrix}\theta(-x)\,e^{\tfrac{\sqrt{1+m_l^2 v_l^4}}{v_l}x},\\
\label{soldefmen-mas}
\psi_-^{(+)}(x)&=&\left[\tfrac{(1+m_r^2 v_r^4)}{v_r^2}\right]^{1/4}\begin{pmatrix}1 \\ i\tfrac{\sqrt{1+m_r^2 v_r^4}}{m_r v_r^2-i}\end{pmatrix}\theta(x)\,e^{-\tfrac{\sqrt{1+m_r^2 v_r^4}}{v_r}x},\\
\label{solskmenmen}
\psi_-^{(-)}(x)&=&\left[\tfrac{(1+m_l^2 v_l^4)}{v_l^2}\right]^{1/4}\begin{pmatrix}1 \\ -i\tfrac{\sqrt{1+m_l^2 v_l^4}}{m_l v_l^2-i}\end{pmatrix}\theta(-x)\,e^{\tfrac{\sqrt{1+m_l^2 v_l^4}}{v_l}x},
\end{eqnarray}
\end{subequations}
where $\theta(x)$ represents the Heaviside step function. Since all the solutions of equations $H^\dag\psi_\pm=\pm i\psi_\pm$ belong to $L^2(\mathbb{R})\otimes\mathbb{C}^2$, the deficiency indices are $(2,2)$ and, the according to Naimark \cite{naimark}, every self-adjoint extensions are parametrized by a $U(2)$ matrix. This matrix defines a unique self-adjoint extension, $H_{U}$, of $H$ with domain characterized by means the set of all functions $\phi\in\mathcal{D}(H^\dag)$ which satisfy the conditions
\begin{equation}\label{conditions}
\begin{pmatrix}
\bar{\psi}_{a2}(0^-) & \bar{\psi}_{a1}(0^-)\\
\bar{\psi}_{b2}(0^-) & \bar{\psi}_{b1}(0^-)
\end{pmatrix}\begin{pmatrix}\phi_{a}(0^{-}) \\ \phi_{b}(0^{-})\end{pmatrix}=\frac{v_r}{v_l}\begin{pmatrix}
\bar{\psi}_{a2}(0^+) & \bar{\psi}_{a1}(0^+)\\
\bar{\psi}_{b2}(0^+) & \bar{\psi}_{b1}(0^+)
\end{pmatrix}\begin{pmatrix}\phi_{a}(0^{+}) \\ \phi_{b}(0^{+})\end{pmatrix}
\end{equation}
where $\psi(0^\pm)\equiv\lim\limits_{x\rightarrow 0^{\pm}}\psi(x)$ and $\phi(0^\pm)\equiv\lim\limits_{x\rightarrow 0^{\pm}}\phi(x)$, and
\begin{align}\label{basemasmen}
\psi_1(x)&= \psi_+^{(+)}(x)+U_{11}\psi_-^{(+)}(x)+U_{21}\psi_-^{(-)}(x)\\
\psi_2(x)&= \psi_+^{(-)}(x)+U_{12}\psi_-^{(+)}(x)+U_{22}\psi_-^{(-)}(x)
\end{align}
where $\psi_1(x),\psi_2(x)\in\mathcal{D}(H_U)$, and $U_{11},U_{12},U_{21},U_{22}$ are complex numbers that determine $U$. The expression
\eqref{conditions} can be written in the form
\begin{equation}\label{boundc1}
\begin{pmatrix}\phi_{a}(0^{+}) \\ \phi_{b}(0^{+})\end{pmatrix}
=\mathbb{T}\begin{pmatrix}\phi_{a}(0^{-}) \\ \phi_{b}(0^{-})\end{pmatrix},
\end{equation}
where the $n_{+}\times n_{-}\;$ matrix $\mathbb{T}$ is given by
\begin{equation}\label{matrixT}
\mathbb{T}=
\frac{v_l}{v_r}
\begin{pmatrix}
\bar{\psi}_{a2}(0^+) & \bar{\psi}_{a1}(0^+)\\
\bar{\psi}_{b2}(0^+) & \bar{\psi}_{b1}(0^+)
\end{pmatrix}^{-1}\begin{pmatrix}
\bar{\psi}_{a2}(0^-) & \bar{\psi}_{a1}(0^-)\\
\bar{\psi}_{b2}(0^-) & \bar{\psi}_{b1}(0^-)
\end{pmatrix},
\end{equation}
whose determinat is given by
\begin{equation}\label{detmatrixT}
\left|\det\mathbb{T}\right|=\frac{v_l}{v_r}.
\end{equation}
The matrix $\mathbb{T}$ gives the matching conditions at the origin. From \eqref{basemasmen}, we can rewrite
the matrix $\mathbb{T}$ in the form
\begin{equation}\label{matrixTu}
\mathbb{T}=\frac{\sqrt{v_l}}{2\bar{U}_{12}\sqrt{v_r}}
\begin{pmatrix}
T_{11}&T_{12}\\
T_{21}&T_{22}
\end{pmatrix}
\end{equation}
with
\begin{subequations}\label{elementosTu}
\begin{align}
T_{11}=&\frac{(1+m_r^2 v_r^4)^\frac{1}{4}((1-i m_l v_l^2)(1+\bar{U}_{11})-(1+i m_l v_l^2)(\bar{U}_{22}+\det(\bar{U})))}{(1+m_l^2 v_l^4)^\frac{1}{4}}\\
T_{12}=&-(1+\det(\bar{U})+\bar{U}_{11}+\bar{U}_{22})((1+m_l^2 v_l^4)(1+m_r^2 v_r^4))^{\frac{1}{4}}\\
T_{21}=&\frac{(i+m_lv_l^2+\bar{U}_{22}(m_l v_l^2-i))(\bar{U}_{11}(m_r v_r^2-i)+m_r v_r^2+i)}{((1+m_l^2 v_l^4)(1+m_r^2 v_r^4))^\frac{1}{4}}\\\nonumber
&-\frac{\bar{U}_{12}\bar{U}_{21}(m_l v_l^2-i)(m_r v_r^2-i)}{((1+m_l^2 v_l^4)(1+m_r^2 v_r^4))^\frac{1}{4}}\\
T_{22}=&\frac{(1+m_l^2 v_l^4)^{\frac{1}{4}}((1-im_r v_r^2)(1+\bar{U}_{22})-(1+im_r v_r^2)(\bar{U}_{11}+\det(\bar{U})))}{(1+m_r^2 v_r^4)^{\frac{1}{4}}}
\end{align}
\end{subequations}

The determinant of \eqref{matrixTu} is given by
\begin{equation}\label{detmatrixTu}
\det\mathbb{T}=\frac{v_l}{v_r}\frac{\bar{U}_{21}}{\bar{U}_{12}}.
\end{equation}
By Comparing \eqref{detmatrixTu} with \eqref{detmatrixT}, we have that $\left|U_{12}\right|=\left|U_{21}\right|$.

\section{Scattering coefficients and the spectra of H}\label{spectra}

In this section we will derive the spectra for the self-adjoint extensions $H_{U}$ from poles of scattering
amplitudes. For this, let us parametrize the unitary matrix $\mathbb{U}$ as
\begin{equation}\label{Uparametrizada}
\mathbb{U}=e^{i\alpha}\mathbb{A},\;
\det(\mathbb{A})=1,
\end{equation}
where
\begin{equation}\label{matrixA}
\mathbb{A}=\begin{pmatrix}
  a_0-ia_3 & -a_2-ia_1 \\
  a_2-ia_1 & a_0+ia_3 \\
\end{pmatrix},
\end{equation}
with $a_0,a_1,a_2,a_3\in\mathbb{R}$, $a_0^2+a_1^2+a_2^2+a_3^2=1$, and $\alpha\in\left[0,\pi\right]$. Notice that the points $\alpha=0$ and
$\alpha=\pi$ have to be identified. Substituting \eqref{Uparametrizada} and \eqref{matrixA} in \eqref{elementosTu}, we obtain the components of matrix $\mathbb{T}$:
\begin{subequations}\label{matrixTa}
\begin{align}
T_{11}=&2ie^{-i\alpha}\frac{(1+m_r^2 v_r^4)^\frac{1}{4}(a_3+\sin\alpha-m_l v_l^2(a_0+\cos\alpha))}{(1+m_l^2 v_l^4)^\frac{1}{4}}\\ 
T_{12}=&2e^{-i\alpha}(a_0+\cos\alpha)((1+m_l^2 v_l^4)(1+m_r^2 v_r^4))^{\frac{1}{4}}\\ 
T_{21}=&2e^{-i\alpha}\Big(\frac{a_0-\cos\alpha+m_l m_r  v_l^2 v_r^2(a_0+\cos\alpha)+m_l v_l^2(a_3-\sin\alpha)-m_r v_r^2(a_3+\sin\alpha)}{((1+m_l^2 v_l^4)(1+m_r^2 v_r^4))^\frac{1}{4}}\Big)\\ 
T_{22}=& -2ie^{-i\alpha}\frac{(1+m_l^2 v_l^4)^{\frac{1}{4}}(a_3-\sin\alpha+m_r v_r^2(a_0+\cos\alpha))}{(1+m_r^2 v_r^4)^{\frac{1}{4}}} 
\end{align}
\end{subequations}

In terms of \eqref{matrixTa}, the matching conditions \eqref{boundc1} are
\begin{equation}\label{bounconda}
\begin{pmatrix}\phi_{a}(0^{+}) \\ \phi_{b}(0^{+})\end{pmatrix}
=\begin{pmatrix}
T_{11}&T_{12}\\
T_{21}&T_{22}
\end{pmatrix}\begin{pmatrix}\phi_{a}(0^{-}) \\ \phi_{b}(0^{-})\end{pmatrix}.
\end{equation}
\indent
Let us assume that an incoming monochromatic wave $\begin{pmatrix}1\\ \sqrt{\frac{E-m_lv_l^2}{E+m_lv_l^2}}\end{pmatrix}e^{ik_{l}x}$, $k_{l}=\frac{\sqrt{E^2-m_l^2v_l^2}}{v_l}$, $E>max (m_l v_l^2,m_r v_r^2)$,
comes from the left, so that the wave function for $x<0$ is $\begin{pmatrix}1\\ \sqrt{\frac{E-m_lv_l^2}{E+m_lv_l^2}}\end{pmatrix}e^{ik_{l}x}+r_l \begin{pmatrix}1\\ -\sqrt{\frac{E-m_lv_l^2}{E+m_lv_l^2}}\end{pmatrix}e^{-ik_{l}x}$, and the wave function
for $x>0$ is $t_l\begin{pmatrix}1\\ \sqrt{\frac{E-m_rv_r^2}{E+m_rv_r^2}}\end{pmatrix}e^{ik_{r}x}$, $k_{r}=\frac{\sqrt{E^2-m_r^2v_r^2}}{v_r}$, $E>max (m_l v_l^2,m_r v_r^2)$, where $r_l$ and $t_l$ are the
reflection and transmission amplitudes, respectively, for an incoming wave come from the left. Then, the matching
conditions \eqref{bounconda} at the origin give
\begin{equation}\label{condRT}
\begin{pmatrix}1+r_l \\*[.4cm] \sqrt{\frac{E-m_lv_l^2}{E+m_lv_l^2}}(1-r_l)\end{pmatrix}=
\mathbb{T}
\begin{pmatrix}t_l \\*[.5cm] \sqrt{\frac{E-m_rv_r^2}{E+m_rv_r^2}}\,t_l\end{pmatrix}
\end{equation}
and then one finally obtains the expressions of $r_l$ and $t_l$ as
\begin{eqnarray}
\label{rl}
&r_l=\frac{\mathfrak{N}}{\mathfrak{D}}\\
\label{tl}
t_l=&2\sqrt{\frac{v_l}{v_r}}\frac{\sqrt{a_1^2+a_2^2}\;e^{-i\tan ^{-1}\left(\frac{a_2}{a_1}\right)}}{\mathfrak{D}}\mathfrak{T}
\end{eqnarray}
with
\begin{multline}\label{n}
\mathfrak{N}=-\sqrt{m_r^2 v_r^4+1} \sqrt{E -m_r v_r^2}\Big(\sqrt{E +m_l v_l^2} (a_3+\sin\alpha-m_l v_l^2(a_0+\cos\alpha))\\
+i(a_0+\cos\alpha)\sqrt{(m_l^2 v_l^4+1) (E-m_l v_l^2)}\Big)
-\sqrt{m_l^2v_l^4+1}\sqrt{E -m_l v_l^2} \sqrt{E +m_r v_r^2}(a_3\\
-\sin\alpha+ m_rv_r^2(a_0+\cos\alpha))-i \sqrt{E +m_l v_l^2}\sqrt{E +m_r v_r^2}(m_l m_r v_l^2 v_r^2(a_0+\cos\alpha)+a_0\\
-\cos\alpha+(m_l v_l^2(a_3-\sin\alpha)-m_r v_r^2 (a_3+\sin\alpha)),
\end{multline}
\begin{multline}\label{d}
\mathfrak{D}=\sqrt{m_r^2 v_r^4+1}\sqrt{E-m_r v_r^2}\Big(\sqrt{E+m_l v_l^2}(a_3+\sin\alpha-m_l v_l^2(a_0-\cos\alpha))\\
-i(a_0+\cos\alpha)\sqrt{(m_l^2 v_l^4+1)(E-m_l v_l^2)}\Big)
-\sqrt{m_l^2v_l^4+1} \sqrt{E -m_l v_l^2} \sqrt{E +m_r v_r^2}(a_3\\
-\sin\alpha+m_rv_r^2(a_0+\cos\alpha))
+i \sqrt{E +m_l v_l^2}\sqrt{E +m_r v_r^2}(m_l m_r v_l^2 v_r^2(a_0+\cos\alpha)+a_0\\
-\cos\alpha)+(m_l v_l^2(a_3-\sin\alpha)-m_r v_r^2 (a_3+\sin\alpha)),
\end{multline}
\begin{equation}\label{te}
\mathfrak{T}=\sqrt[4]{\left(m_l^2 v_l^4+1\right) \left(m_r^2 v_r^4+1\right)} \sqrt{\left(E -m_l
   v_l^2\right) \left(E +m_r v_r^2\right)}
\end{equation}

Since the matrix $\mathbb{T}$ is not real, the transmission amplitudes are different and the self-adjoint extensions are not explicitly time reversal invariant \cite{nogami,coutinho1}.
\indent

Physically, the term $e^{-i\arctan\left(\frac{a_2}{a_1}\right)}$ in \eqref{tl} does not add new information to the phase shift, since that $a_0$, $a_1$, $a_2$ and $a_3$
are independent of the energy, so we can put $a_2 = 0$ without any loss of information. The matrix $\mathbb{T}$ coincides with the corresponding one in the nonrelativistic case \cite{gonzalez-diaz}  precisely when $a_2=0$. In this situation, we have that the matching conditions \eqref{bounconda} are
\begin{equation}\label{bouncondaa20}
\begin{pmatrix}\phi_{a}(0^{+}) \\ \phi_{b}(0^{+})\end{pmatrix}
=\mathbb{T}\Big|_{\scriptscriptstyle a2=0}\begin{pmatrix}\phi_{a}(0^{-}) \\ \phi_{b}(0^{-})\end{pmatrix}.
\end{equation}
whose determinant is
\begin{equation}\label{detmatrixTa20}
\det\mathbb{T}\Big|_{\scriptscriptstyle a2=0}=\frac{v_l}{v_r}
\end{equation}

Making use of $a_0^2+a_1^2+a_3^2=1$, we have $\left|r_l\right|^2+\left|t_l\right|^2\frac{\sqrt{E+m_lv_l^2}\sqrt{E-m_rv_r^2}}{\sqrt{E-m_lv_l^2}\sqrt{E+m_rv_r^2}}\frac{v_r}{v_l}=1$.
The poles of $r_l$ and $t_l$ satisfy the following equation
\begin{equation}\label{poles}
\mathfrak{D}=0.
\end{equation}

The poles of $r_r$ and $t_r$ ($r_r$ and $t_r$ are the reflection and transmission amplitudes, respectively, for incoming wave come from the right) also satisfy \eqref{poles}.The zero values of \eqref{rl} correspond to transmission resonances \cite{newton,taylor}. The zero values of \eqref{tl} are called zero momentum resonances \cite{dombey}, and they occur at $E=\pm m_l v_F^2$ and $E=\pm m_r v_F^2$ ($v_F$ is the Fermi velocity). The anti-particle is described by the hole wave function corresponding to the absence of the state with $E=-m_{l,r}v_F^2$ \cite{dombey}.

In the next subsections, we discuss the spectrum of some self-adjoint extension of \eqref{piecehamiltonian} corresponding to one - dimensional spatial Dirac Hamiltonian: (a) with a equally mixed point interaction potential (PIP) at the origin plus mass jump at the same point, (b) with an inverted mixed PIP at the origin plus mass jump at the same point, (c) with a vector PIP at the origin plus mass jump at the same point, and (d) with a scalar PIP at the origin plus mass jump at the same point. The one - dimensional Dirac Hamiltonian with PIPs without mass jump is analyzed in \cite{dominguez-adame}. In this paper, the selfadjoint extensions of the one - dimensional Dirac operator with point interactions can also be obtained (heuristically) starting from the operator $H=-i\sigma_{x}\frac{d}{dx}+m\sigma_{z}+\mathcal{U}(x)$, with $\mathcal{U}(x)=\left(g_v+\sigma_{z}g_s\right)u(x)$, where $u(x)$ is any peaked function at $x=0$ satisfying $\int_{-\infty}^{\infty}u(x)\,dx=1$. $g_v$ and $g_s$ are the strengths of the vector and scalar components of the potential, respectively. When $g_v=0$, $g_s=0$, $g_v=g_s$ $g_v=-g_s$, we say that we have a scalar point interaction potential, vector point interaction potential, equally mixed point interaction potential and inverted mixed point interaction potential, respectively. For simplicity and comparison, the following sections we will impose that $v_l = v_r = v_F$. Thus, \eqref{detmatrixTa20} equals one, similarly to the case of equal masses.

\subsection{One - dimensional spatial Dirac Hamiltonian  with a equally mixed PIP at the origin plus mass jump at the same point}\label{equallypip}
The boundary conditions corresponding to one - dimensional spatial Dirac Hamiltonian with a vector PIP at the origin plus mass jump at the same point are obtained by the following ids:
\begin{subequations}\label{aes-vectorpip}
\begin{align}
a_0&=-\cos\alpha,\\
a_1&=\sin\alpha,\\
a_3&=0,\\
\cot\alpha&=-\frac{\delta+\left(m_l+m_r\right)v_F^3}{2v_F},\,\delta<0,
\end{align}
\end{subequations}
where $\delta$ is the strength of PIP.
By inserting \eqref{aes-vectorpip} in \eqref{bouncondaa20}, we obtain the matching conditions for this self -adjoint extension:
\begin{equation}\label{bcsequallypip}
\begin{pmatrix}\phi_{a}(0^{+}) \\ \phi_{b}(0^{+})\end{pmatrix}
=\begin{pmatrix}
\frac{\sqrt[4]{m_r^2v_F^4+1}}{\sqrt[4]{m_l^2v_F^4+1}}&0\\
-\frac{i \delta }{v_F \sqrt[4]{m_l^2 v_F^4+1}\sqrt[4]{m_r^2v_F^4+1}}&\frac{\sqrt[4]{m_l^2v_F^4+1}}{\sqrt[4]{m_r^2v_F^4+1}}
\end{pmatrix}\begin{pmatrix}\phi_{a}(0^{-}) \\ \phi_{b}(0^{-})\end{pmatrix}
\end{equation}
\begin{figure}[!ht]
\centerline{\includegraphics[scale=0.35,clip]{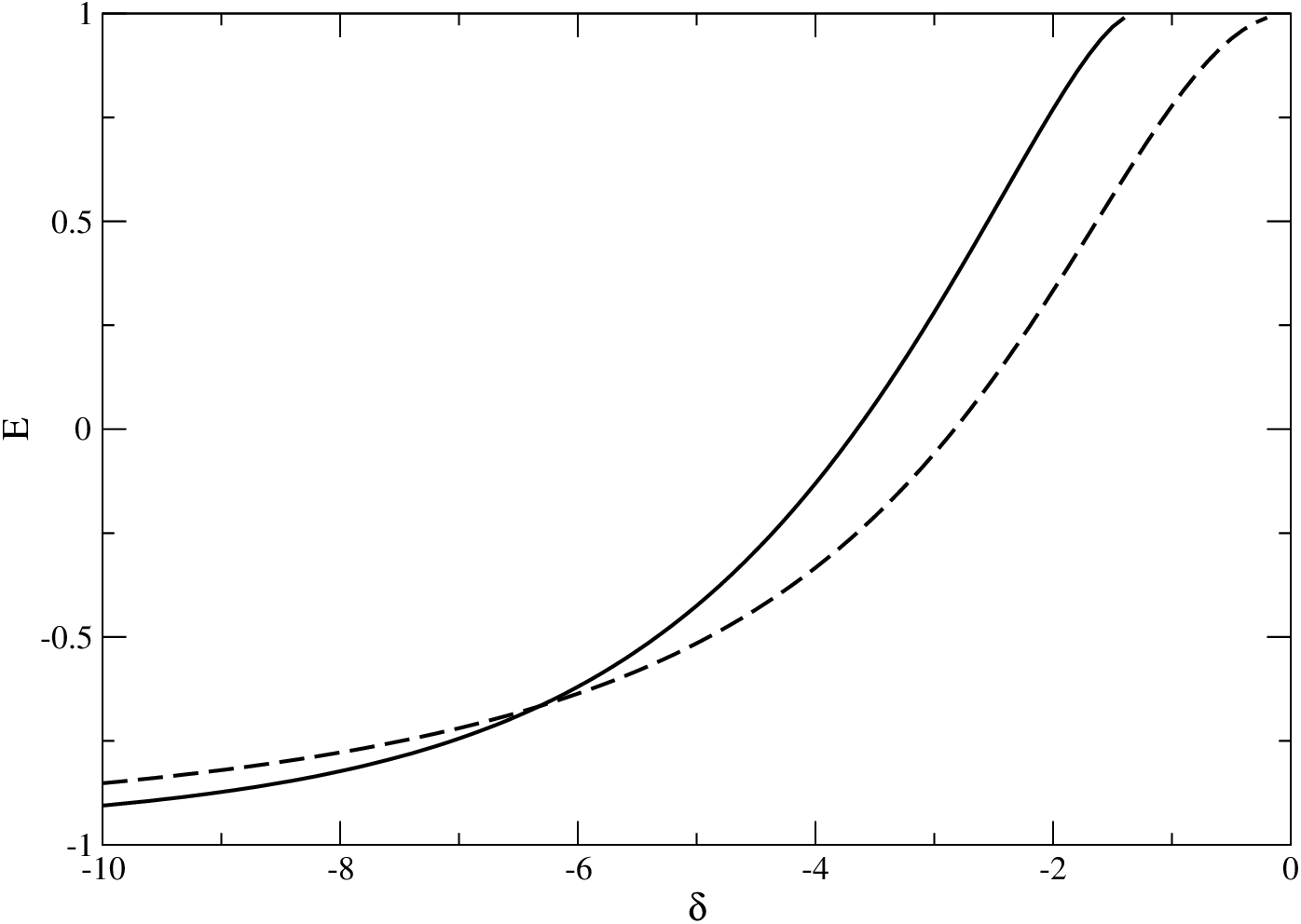}}
\caption{Solution curves of \eqref{spdelta} (solid line) and \eqref{spdeltaaproxsolre} (dashed line) as a function of $\delta$, respectively.}\label{fspdelta}
\end{figure}
By inserting \eqref{aes-vectorpip} in \eqref{poles}, we obtain the spectral equation (bound states energy equation)
\begin{multline}\label{spdelta}
\frac{\delta}{v_F} \sqrt{\left(E +v_F^2 m_l\right)\left(E +v_F^2 m_r\right)}+\sqrt{\left(v_F^4 m_l^2+1\right)\left(v_F^2 m_l-E \right) \left(E +v_F^2 m_r\right)}\\
+\sqrt{\left(v_F^4 m_r^2+1\right) \left(E +v_F^2m_l\right) \left(v_F^2 m_r-E \right)}=0
\end{multline}
The energy of bound states lies between $-min(m_l v_F^2,m_r v_F^2)$ and $min(m_l v_F^2,m_r v_F^2)$. Note that equation \eqref{spdelta} is invariant under the change of $m_l$ by $m_r$.

For $m_l=1,\,m_r=2$ and $v_F=1$, the solution curve of \eqref{spdelta} as a function of $\delta$ is represented in the Figure \ref{fspdelta}. The solution curves intersect at a point, which means that they have the same energy for a given value of $\delta$. The value of the energy is insensitive to ratio of the masses.

As stated in \cite{dominguez-adame}, the boundary of the lower continuum is never reached for finite values of $\delta$ due to the presence of the scalar potential term, but the energy level crosses zero because of the vector potential term (see Figure \ref{fspdelta}).

For $m_l\approx m_r\equiv m$, \eqref{spdelta} becomes
\begin{equation}\label{spdeltaaprox}
2 v_F \sqrt{\left(v_F^4 m^2+1\right) \left(v_F^4 m^2-E^2\right)}+\delta\sqrt{\left(E +v_F^2m\right)^2}=0,
\end{equation}
which gives the value of the energy for bound state
\begin{equation}\label{spdeltaaproxsol}
E=mv_F^2\;\frac{4v_F^2+4m^2v_F^6-\delta^2}{4v_F^2+4m^2v_F^6+\delta^2}
\end{equation}
Defining $\tilde{\delta}=\frac{\delta}{v_F\sqrt{1+m^2v_F^4}}$, the above expression can be rewritten as
\begin{equation}\label{spdeltaaproxsolre}
E=mv_F^2\frac{4-\tilde{\delta}^2}{4+\tilde{\delta}^2}
\end{equation}
The energy \eqref{spdeltaaproxsolre} coincides with the one found in \cite{dominguez-adame} and \cite{albeverio} for the self-adjoint extension called equally mixed potential.

At high energies, we have
\begin{equation}\label{Tlasympdelta}
\left|t_l\right|^2\sim\frac{4 v_F^2 \sqrt{\left(v_F^4 m_l^2+1\right) \left(v_F^4 m_r^2+1\right)}}{v_F^2 \left(\sqrt{v_F^4 m_l^2+1}+\sqrt{v_F^4
   m_r^2+1}\right)^2+\delta ^2}
\end{equation}
so that the transmission does not occur as the potential becomes sufficiently strong. Therefore, the interaction equally mixed PIP at the origin plus mass jump at the same point does confine particles. The same conclusion is reported in \cite{dominguez-adame}.

\subsection{One - dimensional spatial Dirac Hamiltonian with a inverted mixed PIP at the origin plus mass jump at the same point}\label{invertedpip}

The matching conditions for this self -adjoint extension are
\begin{equation}\label{bcsinvertedpip}
\begin{pmatrix}\phi_{a}(0^{+}) \\ \phi_{b}(0^{+})\end{pmatrix}
=\begin{pmatrix}
\frac{(1+m_r^2v_F^4)^\frac{1}{4}}{(1+m_l^2v_F^4)^\frac{1}{4}}&\; -i\,\lambda\, v_F\,((1+m_l^2v_F^4)(1+m_r^2v_F^4))^{\frac{1}{4}}\\
0&\;\frac{(1+m_l^2v_F^4)^\frac{1}{4}}{(1+m_r^2v_F^4)^\frac{1}{4}}
\end{pmatrix}\begin{pmatrix}\phi_{a}(0^{-}) \\ \phi_{b}(0^{-})\end{pmatrix}
\end{equation}
where $\lambda$ is the strength of PIP, $\lambda>0$.
The spectral equation is
\begin{multline}\label{spdeltap}
\sqrt{\left(v_F^4 m_l^2+1\right)\left(v_F^2 m_l-E \right)\left(E +v_F^2
m_r\right)}+\sqrt{\left(v_F^4 m_r^2+1\right)\left(E +v_F^2 m_l\right) \left(v_F^2 m_r-E\right)}\\
-v_F\,\lambda\sqrt{v_F^2m_r-E }\sqrt{v_F^2m_l-E}\sqrt{\left(v_F^4m_l^2+1\right)\left(v_F^4 m_r^2+1\right)}=0,
\end{multline}
where $-min\left(m_l v_F^2,m_r v_F^2\right)<E<min\left(m_l v_F^2,m_r v_F^2\right)$. This equation is invariant under the change of $m_l$ by $m_r$.

For $m_l\approx m_r\equiv m$, \eqref{spdeltap} becomes
\begin{equation}\label{spdeltaaprox}
2\sqrt{\left(v_F^4m^2-E^2\right)}-v_F\,\lambda\sqrt{\left(v_F^4 m^2+1\right)}\left(v_F^2m-E\right)=0,
\end{equation}
which gives the value of the energy for bound state
\begin{equation}\label{spdeltapaproxsolre}
E=-mv_F^2\;\frac{4-\tilde{\lambda}^2}{4+\tilde{\lambda}^2},
\end{equation}
with $\tilde{\lambda}\equiv v_F\,\sqrt{1+m^2v_F^4}\,\lambda$.
The energy \eqref{spdeltapaproxsolre} coincides with the one found in \cite{dominguez-adame} and \cite{albeverio} for the self-adjoint extension called inverted mixed potential.
\begin{figure}[!h]
\centerline{\includegraphics[scale=0.35,clip]{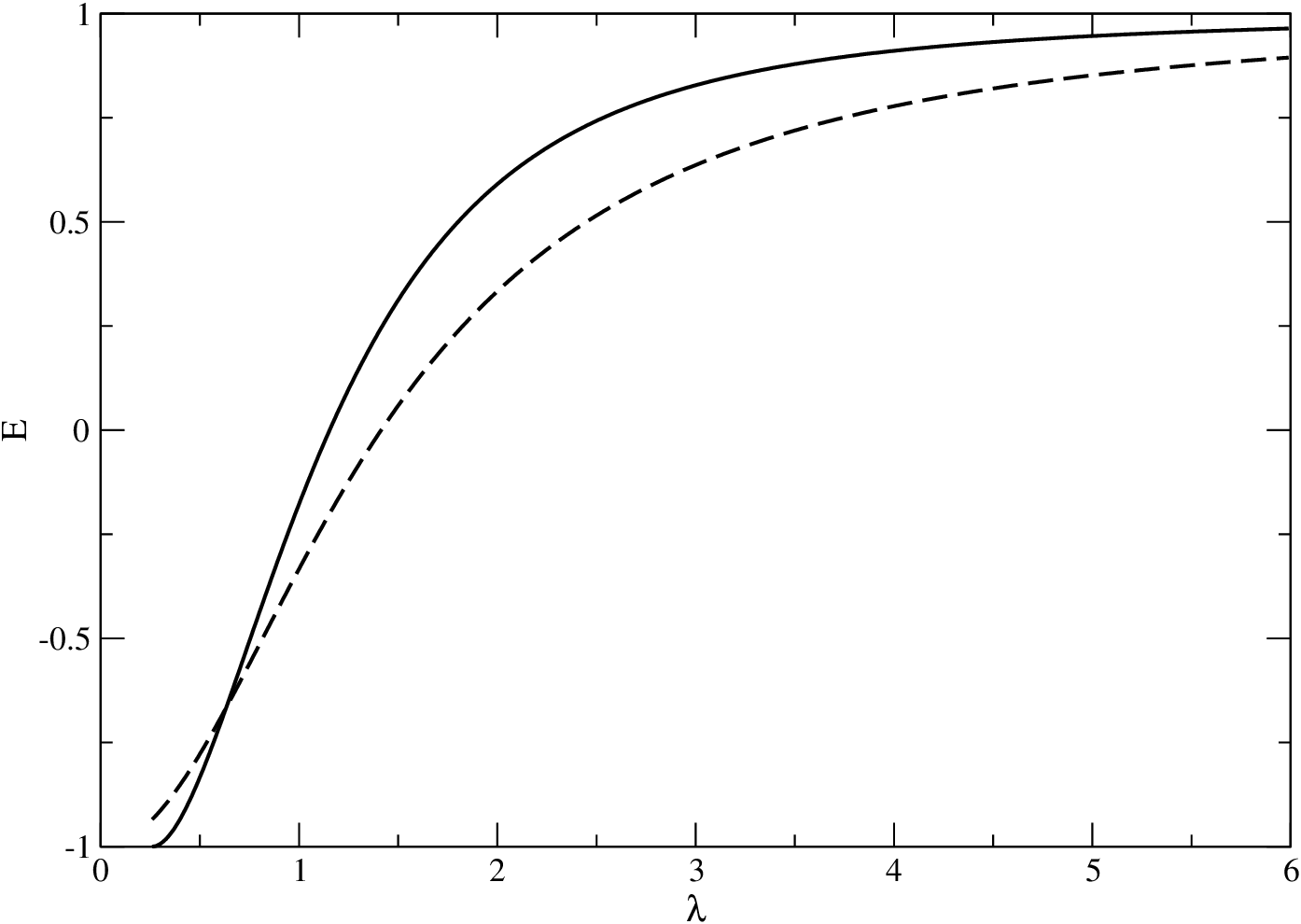}}
\caption{Solution curves of \eqref{spdeltap} (solid line) and \eqref{spdeltapaproxsolre} (dashed line) as a function of $\lambda$, respectively.}\label{fspdeltap}
\end{figure}

For $m_l=1,\,m_r=2$ and $v_F=1$, the solution curve of \eqref{spdeltap} as a function of $\lambda$ is represented in the Figure \ref{fspdeltap}. The solution curves intersect at a point, which means that they have the same energy for $\lambda=0.6324$. For weak coupling, the bound-states energy reaches the lower negative continuum, as distinct from that shown in \cite{dominguez-adame}.

At high energies, the transmission coefficient becomes
\begin{equation}\label{Tlasympdeltap}
\left|t_l\right|^2\sim\frac{4\sqrt{v_F^4 m_l^2+1}\sqrt{v_F^4 m_r^2+1}}{\left(\sqrt{v_F^4 m_l^2+1}+\sqrt{v_F^4
   m_r^2+1}\right)^2+v_F^2\left(v_F^4 m_l^2+1\right)\left(v_F^4
   m_r^2+1\right)\lambda ^2}
\end{equation}
so that the transmission does not occur as the potential becomes sufficiently strong. As in the previous subsection, the interaction inverted mixed PIP at the origin plus mass jump at the same point does confine particles. This same conclusion is reported in \cite{dominguez-adame} for this self-adjoint extension.

\subsection{One - dimensional spatial Dirac Hamiltonian with a pure scalar PIP at the origin plus mass jump at the same point}\label{purescalarpip}
The boundary conditions corresponding to one - dimensional spatial Dirac Hamiltonian with a pure scalar PIP at the
origin plus a mass jump at the same point are
\begin{equation}\label{bcspurescpip}
\begin{pmatrix}\phi_{a}(0^{+}) \\ \phi_{b}(0^{+})\end{pmatrix}
=\begin{pmatrix}
\frac{(1+m_r^2v_F^4)^\frac{1}{4}}{(1+m_l^2v_F^4)^\frac{1}{4}}\cosh{\left(\frac{a}{v_F}\right)}& i\sinh{\left(\frac{a}{v_F}\right)}\\
-i\sinh{\left(\frac{a}{v_F}\right)}&\frac{(1+m_l^2v_F^4)^\frac{1}{4}}{(1+m_r^2v_F^4)^\frac{1}{4}}\cosh{\left(\frac{a}{v_F}\right)}
\end{pmatrix}\begin{pmatrix}\phi_{a}(0^{-}) \\ \phi_{b}(0^{-})\end{pmatrix}
\end{equation}
where $a$ is the strength of PIP, $a<0$.
The spectral equation is
\begin{multline}\label{sppuresc}
\sqrt[4]{v_F^4 m_l^2+1}\sqrt[4]{v_F^4 m_r^2+1}\Big(\sqrt{v_F^2 m_l-E}
\sqrt{v_F^2 m_r-E }\\
+\sqrt{E +v_F^2 m_l}\sqrt{E +v_F^2 m_r}\Big)\sinh\left(\frac{a}{v_F}\right)+\Big(\sqrt{v_F^4 m_l^2+1}\sqrt{(v_F^2 m_l-E)(E +v_F^2 m_r)}\\
+\sqrt{v_F^4m_r^2+1}
\sqrt{(E +v_F^2 m_l)(v_F^2 m_r-E)}\Big)\cosh\left(\frac{a}{v_F}\right)=0,
\end{multline}
where $-min\left(m_l v_F^2,m_r v_F^2\right)<E<min\left(m_l v_F^2,m_r v_F^2\right)$. This equation is invariant under the change of $m_l$ by $m_r$.

\begin{figure}[!h]
\centerline{\includegraphics[scale=0.35,clip]{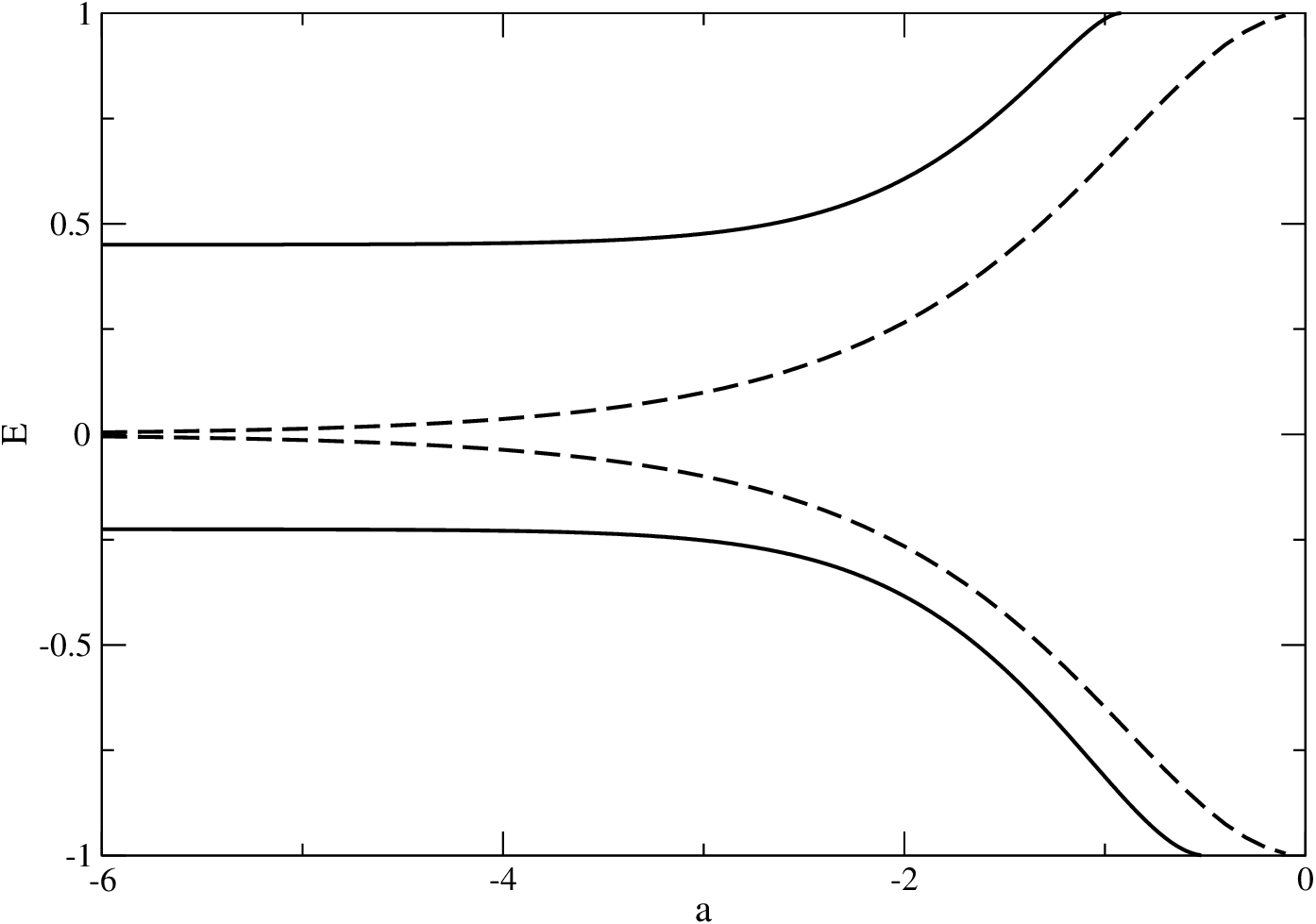}}
\caption{Solution curves of \eqref{sppuresc} (solid line) and \eqref{sppurescpaproxsolre} (dashed line) as a function of $a$, respectively, for $m_l=1$, $m_r=2$ and $v_F=1$.}\label{fsppuresc}
\end{figure}

Pairs of allowed energy values appear, which is a common feature of the other scalar-type potentials \cite{coutinho2}. As seen in Figure \ref{fsppuresc}, the same strength $a$ of the scalar potential can bind particles and antiparticles alike. As stated in \cite{dominguez-adame}, the energy level never reaches zero. The positive and negative energy states remain well separated even if the potential becomes strong.

For $m_l\approx m_r\equiv m$, \eqref{sppuresc} becomes
\begin{equation}\label{sppuresccaprox}
\sqrt{v_F^4m^2-E^2}\cosh\left(\frac{a}{v_F}\right)+v_F^2 m\sinh \left(\frac{a}{v_F}\right)=0,
\end{equation}
which gives the value of the energies for bound states
\begin{equation}\label{sppurescpaproxsolre}
E=\pm\,mv_F^2\,\text{sech}\left(\frac{a}{v_F}\right).
\end{equation}

The energy \eqref{sppurescpaproxsolre} coincides with the one found in \cite{dominguez-adame} for the self-adjoint extension called pure scalar potential. For this potential, at hight energy, the transmission coefficient becomes
\begin{equation}\label{Tlasympuresc}
\left|t_l\right|^2\sim 4\,\frac{\sqrt{(1 + v_F^4 m_l^2)(1 + v_F^4 m_r^2)}\,
  }{\left(\sqrt{1 + v_F^4 m_l^2} + \sqrt{1 +v_F^4 m_r^2}\right)^2}\,\text{sech}^2\left(\frac{a}{v_F}\right)
\end{equation}
Thus, the pure scalar potential leads to particle confinement when $\left|a\right|\rightarrow\infty$ .

\subsection{One - dimensional spatial Dirac Hamiltonian with a pure vector PIP at the origin plus mass jump at the same point}\label{purevectorpip}
The boundary conditions corresponding to one - dimensional spatial Dirac Hamiltonian with a pure vector PIP at the
origin plus a mass jump at the same point are
\begin{equation}\label{bcspurevecpip}
\begin{pmatrix}\phi_{a}(0^{+}) \\ \phi_{b}(0^{+})\end{pmatrix}
=\begin{pmatrix}
\frac{m_r}{m_l}\cos{\left(\frac{a}{v_F}\right)}& -i\sin{\left(\frac{a}{v_F}\right)}\\
-i\sin{\left(\frac{a}{v_F}\right)}&\frac{m_l}{m_r}\cos{\left(\frac{a}{v_F}\right)}
\end{pmatrix}\begin{pmatrix}\phi_{a}(0^{-}) \\ \phi_{b}(0^{-})\end{pmatrix}
\end{equation}
with $a>0$, contrary to the assertion in \cite{dominguez-adame}, where the sign of the strength $a$ is immaterial as far as the existence of bound states.

\begin{figure}[!h]
\centerline{\includegraphics[scale=0.35,clip]{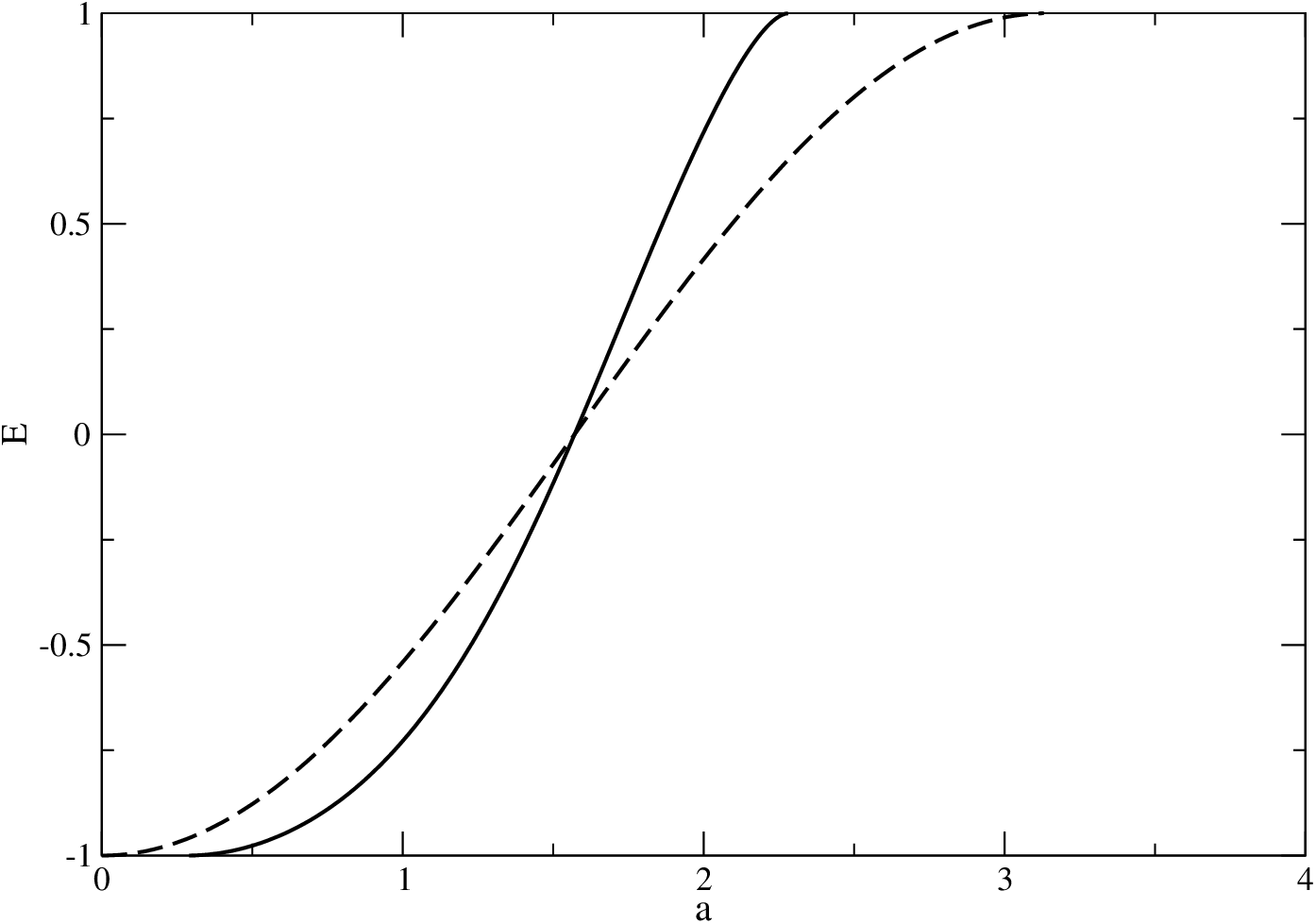}}
\caption{Solution curves of \eqref{sppurevec} (solid line) and \eqref{sppurevecpaproxsolre} (dashed line) as a function of $a$, respectively, for $m_l=1$, $m_r=2$ and $v_F=1$.}\label{fsppurevec}
\end{figure}

The spectral equation is
\begin{multline}\label{sppurevec}
\left(m_l^2 \sqrt{v_F^2 m_r + E} \sqrt{v_F^2 m_l - E} + m_r^2 \sqrt{v_F^2 m_l + E} \sqrt{v_F^2 m_r - E}\right) \cos\left(\frac{a}{v_F}\right)\\
+m_r m_l \Big( \sqrt{v_F^2 m_r + E} \sqrt{v_F^2 m_l + E} -\sqrt{v_F^2 m_l - E} \sqrt{v_F^2 m_r - E}\Big) \sin\left(\frac{a}{v_F}\right)=0,
\end{multline}
where $-min\left(m_l v_F^2,m_r v_F^2\right)<E<min\left(m_l v_F^2,m_r v_F^2\right)$. Unlike the previous cases, \eqref{sppurevec} is not invariant under the change of $m_l$ by $m_r$. For $m_l=1,\,m_r=2$ and $v_F=1$, the solution curve of \eqref{sppurevec} as a function of $a$ is represented in the Figure \ref{fsppurevec}.

For $m_l\approx m_r\equiv m$, \eqref{sppurevec} becomes
\begin{equation}\label{sppureveccaprox}
\sqrt{v_F^4 m^2 - E^2} \cos\left(\frac{a}{v_F}\right)+E\sin\left(\frac{a}{v_F}\right)=0,
\end{equation}
which gives the value of the energies for bound states
\begin{equation}\label{sppurevecpaproxsolre}
E=\pm\,mv_F^2\,\cos\left(\frac{a}{v_F}\right).
\end{equation}
\begin{figure}[!h]
\centerline{\includegraphics[scale=0.35,clip]{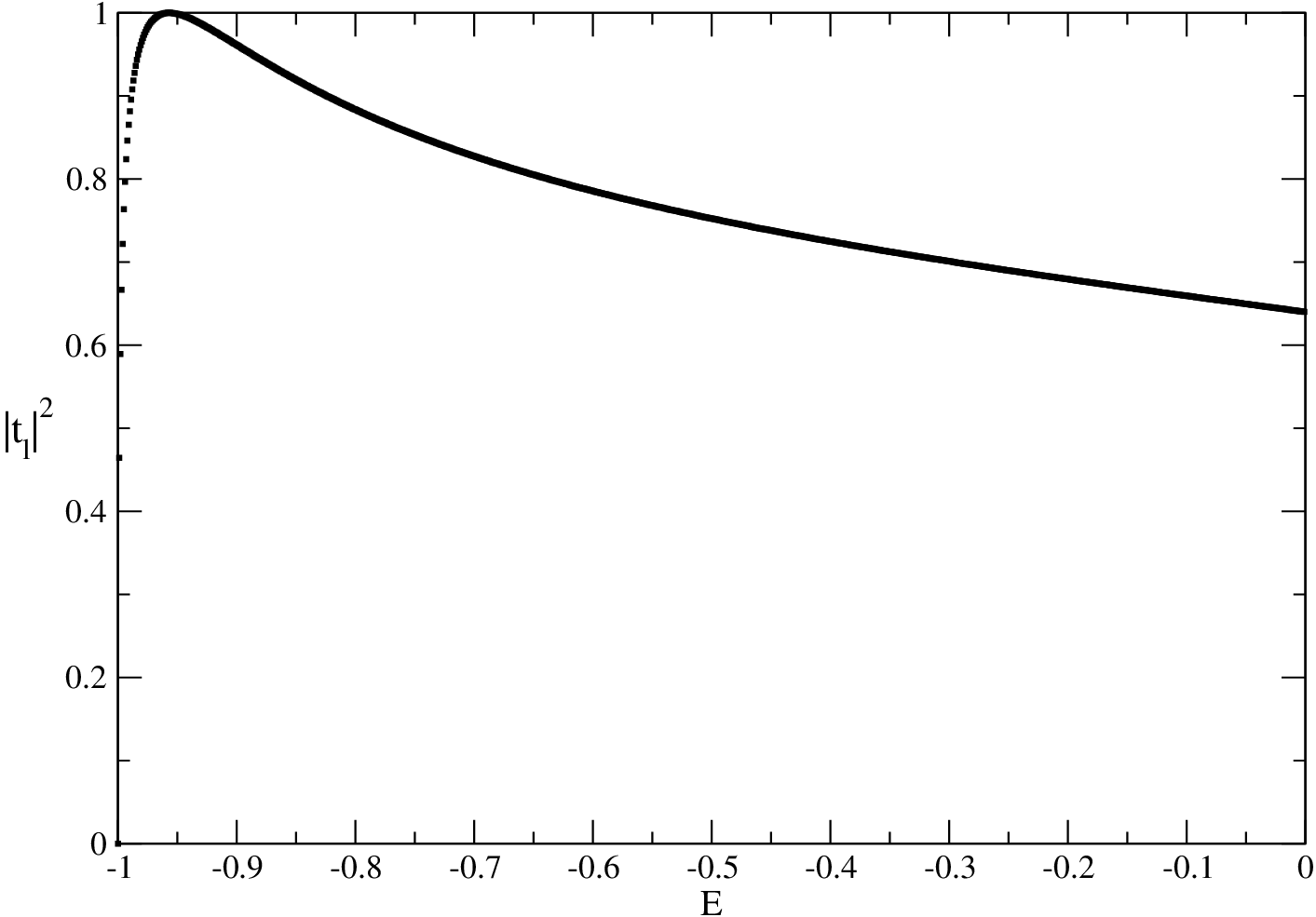}}
\caption{Transmission resonances for $a=\pi$, $m_l=1$, $m_r=2$ and $v_F=1$.}\label{transres1}
\end{figure}

\begin{figure}[!h]
\centerline{\includegraphics[scale=0.35,clip]{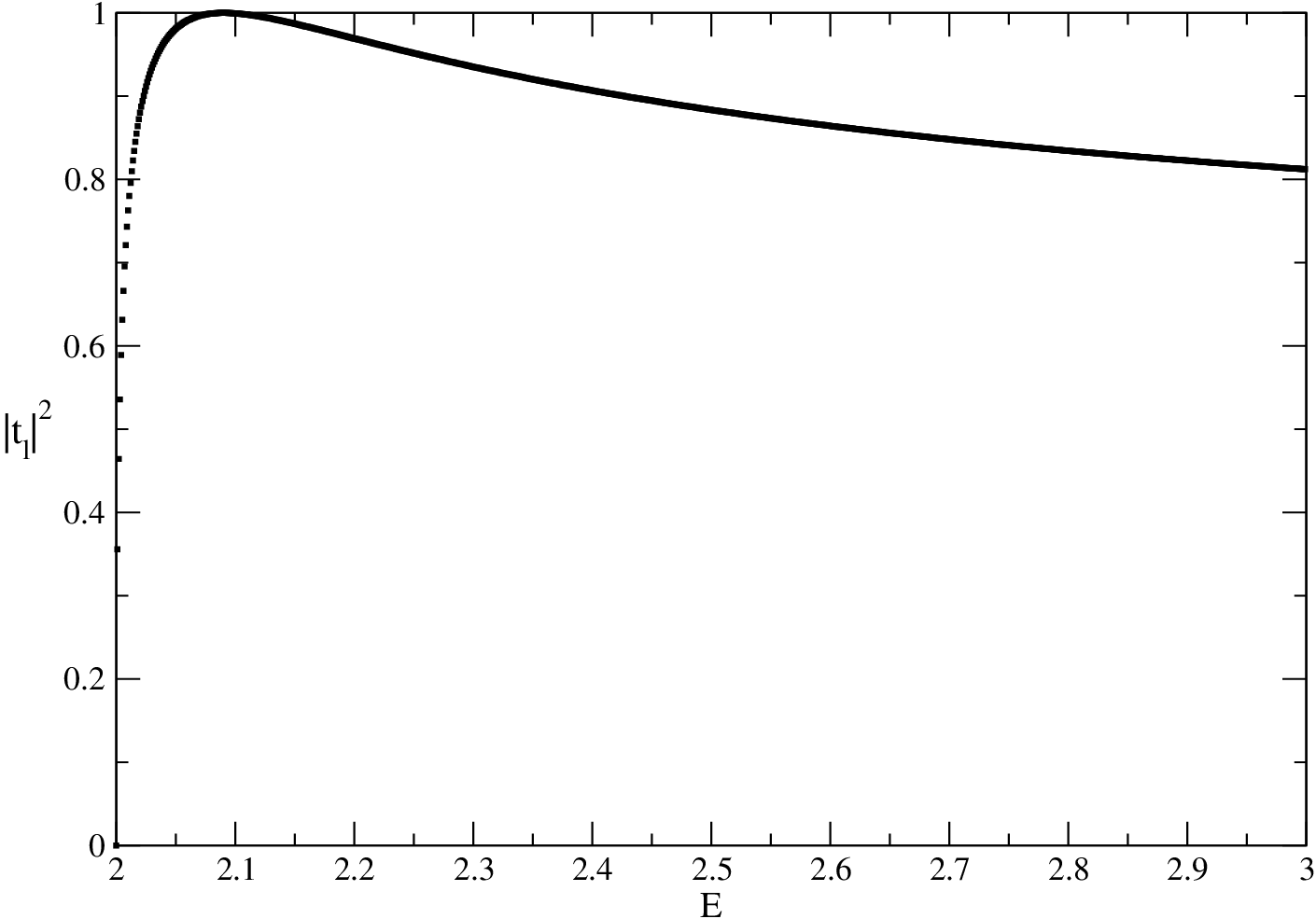}}
\caption{Transmission resonances for $a=\pi$, $m_l=1$, $m_r=2$ and $v_F=1$.}\label{transres2}
\end{figure}

The transmission coefficient is bounded from below,
\begin{equation}\label{Tlpurevec}
\left|t_l\right|^2\geq\frac{8 m_l^2 m_r^2 \sqrt{E -v_F^2 m_l} \sqrt{E +v_F^2 m_l} \sqrt{E -v_F^2 m_r} \sqrt{E
   +v_F^2 m_r}}{4 m_l^2 m_r^2 \sqrt{E -v_F^2 m_l} \sqrt{E +v_F^2 m_l} \sqrt{E -v_F^2
   m_r} \sqrt{E +v_F^2 m_r}+E ^2 \left(m_l^2+m_r^2\right)^2}
\end{equation}
so that the transmission always occurs. There are transmission resonances or virtual bound states \cite{galindo} for values $\frac{a}{v_F}=n\pi$, $n\in\mathbb{Z}$ (see Figures \ref{transres1} and \ref{transres2}).

\section{Concluding remarks}\label{remark}
Using the Von Neumann's theory of self-adjoint extensions and eliminating spurious phases of the transmission amplitude, we found the general matching conditions \eqref{bouncondaa20} that describe each one of the different domains of the various self-adjoint extensions of \eqref{piecehamiltonian}. Using the scattering theory, we obtained the spectrum of each one of the extensions, where each corresponds to a different Hamiltonian operator with interaction.

Finally, we found that of the four different self-adjoint extensions of \eqref{piecehamiltonian}, the first three are confining self-adjoint extensions, while the latter is not. For three of the self-adjoint extensions, there is a value of the strength of the interaction point which the energy of the particle is the same for the mass jump case and without it.

\section{Acknowledgments}
This work was supported by IVIC under Project No. 1089.


\section{References}\label{refs}
\begin{thebibliography}{0}

\bibitem{novoselov1}
K. S. Novoselov, A. K. Geim, S. V. Morozov, D. Jiang, Y. Zhang, S. V. Dubonos, V. I. Grigorieva and A. A. Firsov 2004
{\it Science}
{\bf 306} 666

\bibitem{novoselov2}
K. S. Novoselov,  D. Jiang, T. Booth, V. V. Khotkevich, S. V. Morozov and  A. K. Geim 2005
{\it Proc. Natl. Acad. Sci.}
{\bf 102} 10451

\bibitem{zhu}
S. -L. Zhu,  B. Wang, T. Booth, L. -M. Duan 2007
{\it Phys. Rev. Lett.}
{\bf 98} 260402

\bibitem{ruostekoski}
J. Ruostekoski,  J. Javanainen, and G. V. Dunne, L. -M. Duan 2008
{\it Phys. Rev. A}
{\bf 77} 013603

\bibitem{gomes}
K. K. Gomes,  W. Mar, W. Ko, F. Guinea, and H. C. Manoharan 2012
{\it Nature (London)}
{\bf 483} 306

\bibitem{zhang}
D. -W. Zhang,  L. -B. Shao, Z. -Y. Xue, H. Yan, Z. D. Wang, and S. -L. Zhu 2012
{\it Phys. Rev. A}
{\bf 86} 063616

\bibitem{yannouleas}
C. Yannouleas, I. Romanovsky, and U. Landman 2014
{\it Phys. Rev. B}
{\bf 89} 035432

\bibitem{tamarchenko}
V. I. Tamarchenko and S. A. Ktitorov 1978
{\it Sov. Phys. Solid State}
{\bf 19} 1211

\bibitem{firsova1}
Natalie E. Firsova, Sergey A. Ktitorov, Philip A. Pogorelov 2009
{\it Phys. Lett. A}
{\bf 373} 525

\bibitem{firsova2}
Natalie E. Firsova, Sergey A. Ktitorov 2010
{\it Phys. Lett. A}
{\bf 374} 1270

\bibitem{ktitorov}
S. A. Ktitorov and Natalie E. Firsova 2011
{\it Phys. Solid State}
{\bf 53} 411




\bibitem{jakubsky}
V. Jakubsk\'{y} and D. Krej\v{c}i\v{r}\'{\i}k 2014
{\it Ann. Phys.}
{\bf 349} 268

\bibitem{asorey}
M. Asorey,  A. Ibort, G. Marmo 2005
{\it Int. J. Mod. Phys. A}
{\bf 20} 1001

\bibitem{asorey2}
M. Asorey,  A. Ibort, G. Marmo 2012
{\it Int. J. Geom. Methods Mod. Phys.}
{\bf 9} 1260017

\bibitem{casimir}
H. B. G. Casimir 1948
{\it Proc. K. Ned. Akad. Wet.}
{\bf 51} 793

\bibitem{john}
V. John, G. Jungman, and S. Vaidya 1995
{\it Nucl. Phys. B}
{\bf 455} 505

\bibitem{thouless}
D. J. Thouless, M. Kohmoto, M. P. Nightingale, and M. den Nijs 1982
{\it Phys. Rev. Lett.}
{\bf 49} 405

\bibitem{simon}
M. Reed and B. Simon,
{\it Methods of Modern Mathematical Physics II: Fourier Analysis, Self - Adjointness},
Academic Press Inc., San Diego, California, 1975.

\bibitem{naimark}
M. A. Naimark,
{\it Linear Differential Operators. Vol II},
Frederick Ungar Publishing Company, New York, 1968.

\bibitem{nogami}
Y. Nogami and C. K. Ross 1996
{\it Am. J. Phys.}
{\bf 64}, 923

\bibitem{coutinho1}
F. A. B. Coutinho, Y. Nogami and J. Fernando Perez 1988
{\it J. Phys. A: Math. Gen.}
{\bf 32} L133

\bibitem{gonzalez-diaz}
L. A. Gonz\'{a}lez-D\'{\i}az and S. D\'{\i}az-Sol\'{o}rzano 2013
{\it J. Math. Phys.}
{\bf 54} 042106

\bibitem{newton}
R. G. Newton,
{\it Scattering Theory of Waves and Particles},
McGraw - Hill, Inc., New York, 1966.

\bibitem{taylor}
J. R. Taylor,
{\it Scattering Theory: The Quantum Theory on Nonrelativistic Collisions},
John Wiley \& Sons, Inc., New York, 1972.

\bibitem{dombey}
N. Dombey, P. Kennedy and A. Calogeracos 2000
{\it Phys Rev Lett.}
{\bf 85} 1787

\bibitem{dominguez-adame}
F. Dom\'{\i}nguez - Adame and E. Maci\'{a} 1989
{\it J. Phys. A}
{\bf 22} L419

\bibitem{albeverio}
S. Albeverio, F. Gesztesy, R. H{\o}egh-Krohn and H. Holden,
{\it Solvable Models in Quantum Mechanics},
Second Edition, AMS, 2005.

\bibitem{coutinho2}
F. A. B. Coutinho, Y. Nogami and F. M. Toyama 1988
{\it Am. J. Phys.}
{\bf 56} 904

\bibitem{galindo}
A. Galindo and P. Pascual,
{\it Mec\'{a}nica Cu\'{a}ntica},
Editorial Alhambra, Madrid. 1978.
\end{thebibliography}
\end{document}